\documentclass[pdflatex,sn-nature]{sn-jnl}

\usepackage{graphicx}%
\usepackage{multirow}%
\usepackage{amsmath,amssymb,amsfonts}%
\usepackage{amsthm}%
\usepackage{mathrsfs}%
\usepackage[title]{appendix}%
\usepackage{xcolor}%
\usepackage{textcomp}%
\usepackage{manyfoot}%
\usepackage{booktabs}%
\usepackage{algorithm}%
\usepackage{algorithmicx}%
\usepackage{algpseudocode}%
\usepackage{listings}%
\usepackage{braket}
\usepackage{siunitx}
\theoremstyle{thmstyleone}%

\theoremstyle{thmstyletwo}%

\theoremstyle{thmstylethree}%

\begin{document}

\title[Article Title]{Quantum-optimal nonlinear microscopy with classical light}


\author[1]{\fnm{Joshua L.} \sur{Reynolds}}

\author[2]{\fnm{Shaun C.} \sur{Burd}}

\author[2]{\fnm{Tzu-Chieh} \sur{Yen}}

\author[3,4]{\fnm{Samsuzzoha} \sur{Mondal}}

\author[3,4]{\fnm{Soichi} \sur{Wakatsuki}}

\author*[1,2]{\fnm{Mark A.} \sur{Kasevich}}\email{kasevich@stanford.edu}

\affil[1]{\orgdiv{Department of Applied Physics}, \orgname{Stanford University}, \orgaddress{\city{Stanford}, \state{CA} \postcode{94305}, \country{USA}}}

\affil[2]{\orgdiv{Department of Physics}, \orgname{Stanford University}, \orgaddress{\city{Stanford}, \state{CA} \postcode{94305}, \country{USA}}}

\affil[3]{\orgdiv{Department of Structural Biology}, \orgname{Stanford University}, \orgaddress{\city{Stanford}, \state{CA} \postcode{94305}, \country{USA}}}

\affil[4]{\orgdiv{Biological Sciences Division}, \orgname{SLAC National Accelerator Laboratory}, \city{Menlo Park}, \state{CA} \postcode{94025}, \country{USA}}


\abstract{

Nonlinear optical processes are used in biological microscopy to surpass the diffraction limit on resolution \cite{Hell1994}, image deeper into brain tissues \cite{Oheim2001, Horton2013}, and identify biomolecules without exogenous labels \cite{Zumbusch1999, Freudiger2008}. These techniques typically require high optical intensities to increase the strength of the nonlinear interactions, which can perturb native biochemistry and damage or kill living samples. Stimulated Raman scattering (SRS) microscopy visualizes the spatial distribution of molecules using a nonlinear interaction between light and chemically specific molecular vibrations \cite{Prince2017, Shen2019, Manifold2022, Min2025}. However, the detection of biomolecules at low concentrations is limited by the total photon dose that can be applied before photodamage alters the sample, and photon shot noise sets the minimum achievable noise floor for most microscopes. Here we demonstrate a cavity-enhanced SRS microscope that is more sensitive than an equivalent conventional SRS microscope by up to \qty{8.3(7)}{\decibel} in spectroscopy and \qty{8.6(1)}{\decibel} in cell imaging. These results approach quantum limits on sensitivity and demonstrate that quantum states of light are sufficient but not necessary to enhance the sensitivity of microscopy techniques that are limited by photodamage.

}

\maketitle

\newpage 

%\section{Main}\label{sec1}

Measurement precision in optical microscopy is fundamentally limited by the quantum uncertainty inherent in optical fields. For coherent states, which are the quantum mechanical representations of classical light fields, this uncertainty manifests as photon shot noise, where measurements detecting on average $N$ photons will exhibit fluctuations with a standard deviation of $\sqrt{N}$. Increasing the number of photons improves the measurement signal-to-noise ratio (SNR). However, the onset of photostress or photodamage will limit the total useful photon dose that can be applied to a sample, and it is a key challenge in microscopy to maximize the amount of information that is obtained from each applied photon.

In stimulated Raman scattering (SRS) microscopy, two intense laser beams---the higher-frequency pump beam and the lower-frequency Stokes beam---drive transitions between molecular vibrational states of frequency $\omega = \omega_P - \omega_S$, where $\omega_{S(P)}$ is the optical frequency of the Stokes (pump) beam. The molecular concentration near the optical focus is linearly related to the number of photons scattered from the pump beam into the Stokes beam $N_\text{SRS} \propto C N_S N_P$, where $C$ is the molecular concentration and $N_{S(P)}$ is the Stokes (pump) photon dose \cite{Min2024a}. Images are formed by detecting changes in the optical intensities as a function of position, with a shot-noise-limited SNR that scales as $\sqrt{N_S}$ ($\sqrt{N_P}$) for measurements of stimulated Raman gain (loss) on the Stokes (pump) beam. With \unit{\per\centi\meter}-scale spectral resolution, SRS microscopy can be used to distinguish biomolecules based on their vibrational spectra and measure their concentrations with spatial resolutions of as low as \qty{86}{\nano\meter} \cite{Lin2025}. 

In comparison to fluorescence microscopy, SRS microscopy provides label-free image contrast for molecules with distinct vibrational spectra and is not subject to photobleaching \cite{Freudiger2008, Cheng2015, Lu2015, Oh2022}. Furthermore, vibrational probes can be more highly multiplexed in the spectral domain than fluorophores \cite{Wei2017, Hu2018, Shi2022, Shou2021}. However, most SRS microscopes are limited in sensitivity to millimolar concentrations, and many extensions of SRS that are being explored to push the sensitivity into the nanomolar and single-molecule regimes suffer from increased experimental complexity, invasiveness, or background signals \cite{Shi2018, Xiong2019, Zong2019, Yu2024, Lin2025}. For all SRS techniques, the measurement sensitivity rises with increasing photon dose, but, when the dose is limited by photodamage \cite{Fu2006,Nan2006,Zhang2011,Casacio2021,Zhang2022,Sarri2024, Lin2025}, the achievable concentration sensitivity is bounded due to photon shot noise.

Quantum states of light can increase the measurement SNR beyond the shot-noise limit for a given optical dose but are challenging to generate and use in high-performance microscopes. SRS microscope sensitivity has been improved by using squeezed states of light, which exhibit intensity fluctuations below the shot-noise limit \cite{Andrade2020, Casacio2021, Xu2022a, Xu2022, Xu2023, Terrasson2024, Xu2025, Akatev2025}. The highest reported SNR enhancements of \qty{3.6}{\decibel} require slow imaging speeds and fixed samples \cite{Andrade2020, Akatev2025}. Imaging of live biological samples with optical intensities and imaging speeds competitive with conventional state-of-the-art microscopes has been demonstrated with SNR enhancements of up to \qty{1.3}{\decibel} \cite{Casacio2021, Terrasson2024}. 

The sequential interrogation of a sample with the same coherent state also increases the measurement SNR for a given optical dose, demonstrating that non-classical states are merely sufficient to increase the dose-limited SNR \cite{Luis2002,Giovannetti2006}. For example, $m$ sequential applications of the same $N/m$ photons applies a fixed total dose $N$, and generates, for weak scatterers, a fixed total signal that is proportional to $N$ and independent of $m$. However, the detected optical power is reduced by $m$ and the associated shot noise by $\sqrt{m}$, increasing the SNR at constant total dose relative to a single-pass ($m = 1$) measurement. 

Optical cavities enable these repeated probe-sample interactions. Cavity-enhanced spectroscopy provides extreme sensitivities for gas-phase samples, for which imaging is unnecessary \cite{Yang2023, Liang2025}. Cavity-enhanced microscopes combine this sensitivity with micron- and sub-micron spatial resolutions and have been implemented with many imaging modalities \cite{Mader2015, Juffmann2016, Israel2023, Cuevas2023, Lueghamer2025, Pittrich2025}, including spontaneous Raman microscopy, where a Fabry-Perot microcavity was used to enhance the rate of spontaneous Raman scattering for label-free imaging \cite{Huemmer2016}. Here, we demonstrate a cavity-enhanced SRS microscope that provides enhancements of up to \qty{8.3(7)}{\decibel} in spectroscopy of bulk solvents and \qty{8.6(1)}{\decibel} in cell imaging over equivalent single-pass methods. We show that this method nearly saturates quantum limits for SRS microscope sensitivity. 

\begin{figure}
    \centering
    \makebox[\textwidth]{\includegraphics[width=18cm]{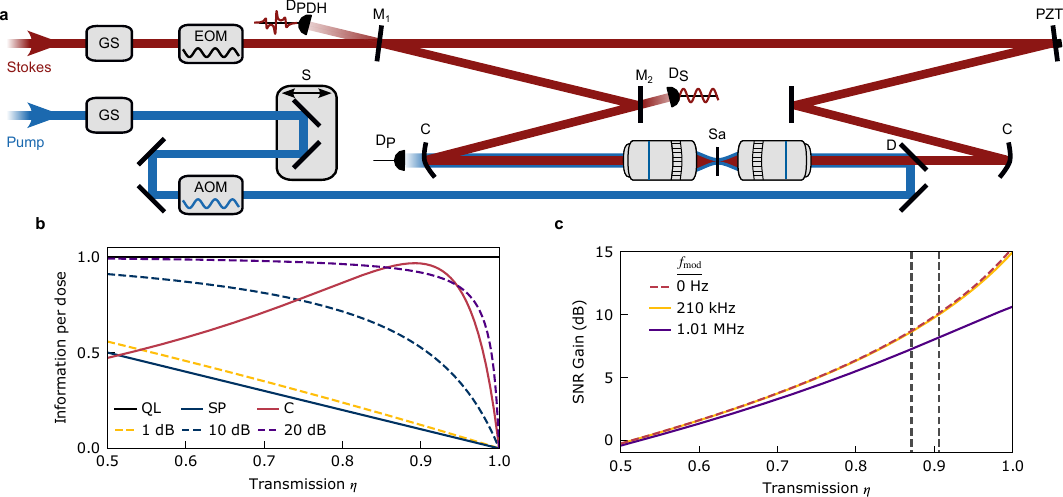}}
    \caption{Cavity-enhanced SRS microscopy. (a) Stokes pulses (red) are enhanced in an optical cavity and are spatially and temporally overlapped with pump pulses (blue) at the cavity waist formed between two microscope objectives to drive SRS. The Stokes (pump) beam center wavelength is \qty{1045}{\nano\meter} (\qty{800}{\nano\meter}). Both pulses are temporally stretched to durations of \qty{2.5}{\pico\second} with optical gratings. The cavity $f_\text{FSR}$ is stabilized to $f_\text{rep} \sim $ \qty{80}{\mega\hertz}. GS, grating stretcher; EOM, electro-optic modulator; AOM, acousto-optic modulator for amplitude modulation at $f_\text{mod}$; S, translation stage; PZT, mirror mounted on piezoelectric transducer; M$_{1(2)}$, input (output) coupler with intensity transmission $T_{1(2)}$; D, dichroic mirror; C, concave mirror; Sa, sample; D$_\text{P}$, pump detector; D$_\text{PDH}$, PDH detector; D$_\text{S}$, Stokes detector. (b) Information per average dose for SRS signal estimation with round-trip transmission $\eta$, normalized to the quantum limit (black; QL), for classical single-pass (blue; SP); 1, 10, and \qty{20}{\decibel} squeezed state (orange, blue, purple; dashed; SQ); and cavity-enhanced ($T_1 = 0.004$ and $T_2 = 0.104$) (red; C) measurements. (c) Theoretical SNR gain (red; dashed) over an ESM for the cavity-enhanced SRS microscope. High-frequency modulation signals are attenuated by the cavity, as shown for $f_\text{mod}$ = \qty{210}{\kilo\hertz} (orange) and \qty{1.01}{\mega\hertz} (purple). Dashed gray lines are the transmissions measured in this work ($\eta = 0.871$, 0.906). See Methods for details.}
    \label{fig:setup}
\end{figure}

The Fisher information per average photon dose $\tilde{\mathcal{F}}$ bounds from below the variance in an estimate of the SRS signal for a given measurement strategy, and thus provides a quantitative comparison of quantum and classical measurements (Fig. \ref{fig:setup}b, see Methods). A quantum-optimal measurement saturates the limit on $\tilde{\mathcal{F}}$ that is calculated by maximizing $\tilde{\mathcal{F}}$ over all possible quantum states and estimation strategies.

Interrogating the sample with Fock state probes in a single-pass configuration is one quantum-optimal measurement strategy, which can be asymptotically approximated using increasingly squeezed states of light \cite{Adesso2009, Whittaker2017, Allen2020}. This strategy yields a maximum increase in information over single-pass measurements that use coherent state probes of $1/(1-\eta \chi)$, where $1-\eta < 1$ is the optical intensity loss and $\chi$ is the stimulated Raman intensity loss or gain, which typically differs from 1 by less than $10^{-4}$. Cavity-enhanced intensity measurements using coherent states saturate the quantum limit on the Fisher information per average number of input photons \cite{Belsley2022} and per average number of intracavity photons \cite{Tao2024} but have not been shown to be quantum-optimal per average sample dose. We show that optimal cavity-enhanced measurements saturate the quantum limit on $\tilde{\mathcal{F}}$ when estimating the SRS signal. 

\begin{figure}[h]
    \centering
    \makebox[\textwidth]{\includegraphics[width=18cm]{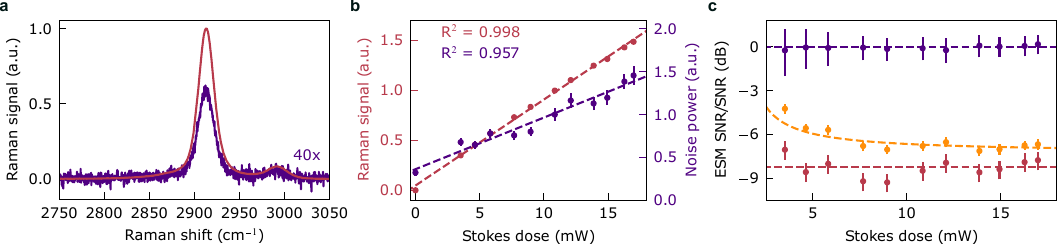}}
    \caption{Cavity-enhanced SRS spectroscopy of DMSO. (a) Cavity-enhanced (red) and single-pass (purple) spectra of DMSO measured with the same input power at $M_1$ (Fig. \ref{fig:setup}). The Stokes dose is \qty{11}{\milli\watt} for the cavity-enhanced measurement. The measured single-pass signal is magnified by 40$\times$ for clarity. (b) Cavity-enhanced SRS signal amplitude at \qty{2913}{\per\centi\meter} (red) and noise power (purple) as a function of the Stokes dose, normalized to measured values for an \qty{11}{\milli\watt} Stokes dose. Error bars are 95\% confidence intervals. Dashed lines are linear fits. (c) ESM SNR (purple) compared to the cavity-enhanced SNRs with (orange) and without (red) electronic noise, normalized to the ESM limit for each Stokes dose. Error bars are 95\% confidence intervals on the measured SNRs. Dashed lines guide the eye.}
    \label{fig:spectroscopy}
\end{figure}

An optical cavity with a free spectral range $f_\text{FSR}$ actively stabilized to match the repetition rate $f_\text{rep} \sim $ \qty{80}{\mega\hertz} of a femtosecond laser is used to resonantly enhance the Stokes beam (Fig. \ref{fig:setup}a). Two microscope objectives inside the optical cavity generate a cavity waist with a 1/$e^2$ intensity radius of \qty{1.4(2)}{\micro\meter} for SRS microscopy (see Methods, Extended Data Fig. \ref{fig:waist}). The sample is scanned through the cavity waist to generate images.

The center wavelength of the Stokes beam is \qty{1045}{\nano\meter}, and the pump beam is tuned between \qty{795}{\nano\meter} and \qty{805}{\nano\meter} to probe Raman transitions in the C-H stretch region around \qty{3000}{\per\centi\meter} for label-free biological imaging. The Stokes and pump pulses are temporally stretched to durations of about \qty{2.5}{\pico\second} so that different Raman shifts can be probed by adjusting the pump--Stokes delay time \cite{Andresen2011}. On every round-trip, the circulating Stokes pulse is overlapped with a new pump pulse in the sample to drive SRS. The pump beam is amplitude modulated for shot-noise-limited lock-in detection of the stimulated Raman gain signal on the Stokes light that leaks out of the cavity. The modulation frequency $f_\text{mod}$ (typically \qty{210}{\kilo\hertz} or \qty{1.01}{\mega\hertz}, see Methods) is chosen to avoid technical noise. 

For weak SRS signals, cavity input and output coupling mirrors $M_1$ and $M_2$ (Fig. \ref{fig:setup}a) with intensity transmissions $T_1$ and $T_2$, respectively, and a round-trip intensity loss $1-\eta$, the anticipated gain $G$ in the shot-noise-limited SNR---defined as the ratio of the signal amplitude to the noise amplitude---relative to an equivalent single-pass measurement conducted with the same average dose (ESM) (see Methods), is
\begin{equation}
    G = \frac{\sqrt{T_2}}{1-\sqrt{(1-T_1)(1-T_2)\eta \chi}},
    \label{gain}
\end{equation}
which is maximized by minimizing $T_1$ and setting $1-T_2 = (1-T_1)\eta\chi$. The increase in information relative to an ESM is $G^2$. For the optimal settings of $T_1$ and $T_2$, $G^2 \rightarrow 1/(1-\eta \chi)$, demonstrating that cavity-enhanced measurements saturate the quantum limit on information per dose and cannot be outperformed under equivalent measurement conditions by using non-classical states of light (see Methods, Supplementary Material A). In the following experiments, $T_1 = 0.004$, and $T_2 = 0.104 \approx 1-(1-T_1)\eta$, as required for near-quantum-optimal performance (Fig. \ref{fig:setup}b). 

The theoretical metrological gain for the cavity-enhanced microscope is shown in Fig. \ref{fig:setup}c. The measured $\eta$ are limited by losses in the sample and microscope objectives, which is typical for biological imaging. In this regime, quantum-optimal measurements can be performed with low-finesse optical cavities. However, the use of high-frequency modulation for lock-in detection is still limited by the finite bandwidth of the dynamic cavity response. Because the Raman signal is measured as the heterodyne beat note between the carrier and the amplitude-modulation sidebands \cite{Min2011}, the SRS signal is reduced by the cavity attenuation of the sideband amplitudes \cite{Yariv2000}, as shown quantitatively in Fig. \ref{fig:setup}c. Nevertheless, the reduction in SRS measurement bandwidth and thus imaging speed is negligible. Typical integration times in SRS microscopy are at least \qty{1}{\micro\second} \cite{Freudiger2008, Casacio2021, Shi2022, Yu2024}, which is far longer than the cavity lifetime ($\sim 60$ \unit{\nano\second}). 

Example SRS spectra of the organic solvent dimethyl sulfoxide (DMSO) measured with the cavity-enhanced microscope are shown in Fig. \ref{fig:spectroscopy}a. The cavity finesse---defined as the ratio of $f_\text{FSR}$ to the full width at half-maximum (FWHM) of the cavity resonance (Extended Data Fig. \ref{fig:waist})---is 29.6(7) and determines $\eta$ = 0.906(4) (see Methods). The amplitude modulation frequency of the pump beam is \qty{210}{\kilo\hertz}, and the expected SNR attenuation due to the dynamic response of the cavity is $<$\qty{0.1}{\decibel} (Fig. \ref{fig:setup}c). 

The cavity-enhanced SRS signal and the optical noise power grow linearly with increasing Stokes dose, as expected for shot-noise-limited optical measurements (Fig. \ref{fig:spectroscopy}b). To estimate the performance of an ESM, two mirrors are inserted to pick off and detect the Stokes beam after a single beam--sample interaction (see Methods). For the same input optical power at $M_1$, the single-pass signals are 69(4)$\times$ weaker (for example, Fig. \ref{fig:spectroscopy}a). The dose is 79(2)$\times$ lower, and the power incident on $D_S$ is 7.9(2)$\times$ lower. 

The expected ESM signal is estimated by scaling the measured single-pass signal by the ratio of the cavity-enhanced dose to the measured single-pass dose. The ESM noise is modeled as pure shot noise, which sets the minimum achievable noise floor, based on the measured cavity-enhanced noise powers (see Methods for details). The measured change in SNR in dB, $20$ $ \text{log}_{10}(\text{ESM SNR}/\text{SNR})$ reaches up to \qty{-7.2(5)}{\decibel} at a dose of \qty{14}{\milli\watt} (Fig. \ref{fig:spectroscopy}c). The measured enhancement is limited by technical noise from the photodiode, which is most relevant at low doses. After subtracting this technical noise to estimate the shot-noise-limited performance, the average enhancement is \qty{8.3(7)}{\decibel}. The uncertainties are standard errors and are dominated by uncertainties in the extrapolated shot noise for the ESMs. 

The quantum limit on SNR enhancement for a transmission $\eta = 0.906$ is \qty{10.3}{\decibel}. In the absence of electronic noise, these measurements come within \qty{2.0(7)}{\decibel} of this limit on sensitivity enhancement for the given microscope parameters. An equivalent \qty{8.3}{\decibel} sensitivity enhancement in a single-pass measurement would require an input optical field squeezed by \qty{12.2}{\decibel} due to the loss-induced degradation of the squeezing, assuming perfect detector quantum efficiency. Quantum limits on sensitivity including detector inefficiencies are discussed in Supplementary Material B. 

\begin{figure}
    \centering
    \makebox[\textwidth]{\includegraphics[width=18cm]{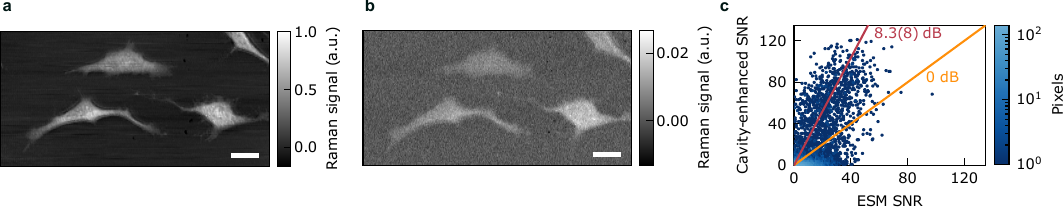}}
    \caption{Cavity-enhanced SRS imaging. HeLa cells imaged under (a) cavity-enhanced and (b) single-pass conditions for the same input power at $M_1$.  Pixel values correspond to relative signal amplitudes. Scale bars are \qty{20}{\micro\meter}. (c) Scatter plot of the cavity-enhanced and ESM SNRs for each pixel, calculated after down-sampling each image onto a \qtyproduct{1.5 x 1.5}{\micro\meter} grid. The SNR enhancement is \qty{8.3(8)}{\decibel} (red) relative to an ESM (orange).}
    \label{fig:imaging}
\end{figure}

Cavity-enhanced and single-pass SRS microscopy of dose-sensitive samples is compared by imaging HeLa mammalian cells at a Raman shift of \qty{2913}{\per\centi\meter} (Fig. \ref{fig:imaging}a-b). High optical doses could be used to damage cells, as evidenced by burning and membrane blebbing, which was observed with cavity-enhanced SRS microscopy (see, for example, Extended Data Fig. \ref{fig:blebbing}). The illumination conditions sufficient to cause damage were variable, but many mammalian cell samples were sufficiently stable for repeated imaging. 

Each \qtyproduct{210 x 105}{\micro\meter} image in Fig. \ref{fig:imaging} was acquired in about 5 minutes. The cavity finesse was 26(1) [$\eta = 0.880(8)$] when the waist was off of a cell, and 24.9(7) [$\eta = 0.871(6)$] when the waist was on a cell, indicating that further enhancements could be achieved by reducing off-cell round-trip losses. The quantum-optimal SNR enhancement of a measurement with $\eta = 0.871$ would be \qty{8.9}{\decibel} over an ESM.

The measured signal amplitudes contain the SRS signal and a sample-independent background, which is removed using a linear baseline correction (see Methods). These background signals are attributed to cross-phase modulation at the optical focus \cite{Genchi2023}. To estimate the enhancement, the images are downsampled, and the single-pass and cavity-enhanced SNRs are compared for each resulting pixel (Fig. \ref{fig:imaging}c). The cavity-enhanced SNR is increased by \qty{8.3(8)}{\decibel} and \qty{8.6(1)}{\decibel} in the absence of an electronic noise floor relative to an ESM, approaching the \qty{8.9}{\decibel} quantum limit on enhancement for $\eta = 0.871$. This enhancement improves measurements of subcellular biomolecular concentration gradients and enables detection of weak SRS signals, for example, from cell edges and cellular projections, which could otherwise go undetected in an ESM. The same \qty{8.6}{\decibel} SNR enhancement would require input light squeezed by \qty{19.8}{\decibel} at $\eta = 0.871$. 

\begin{figure}
    \centering
    \includegraphics[width=9cm]{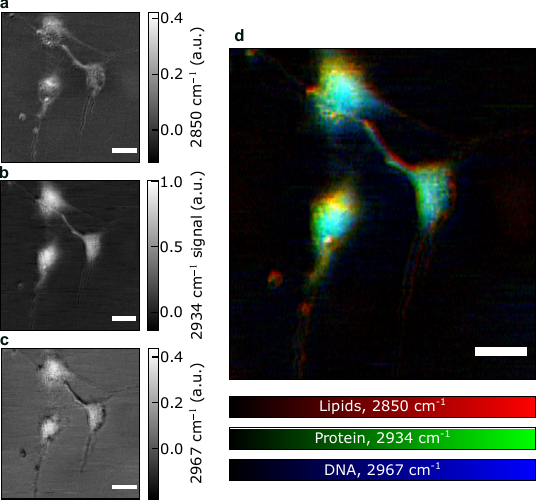}
    \caption{Cavity-enhanced hyperspectral imaging. Cavity-enhanced SRS images acquired sequentially for Raman shifts of (a) 2850, (b) 2934, and (c) and 2967 \unit{\per\centi\meter}, corresponding to Raman bands associated with lipids, protein, and DNA, respectively. Pixel values correspond to relative signal amplitudes. (d) Hyperspectral image combining the measurements for lipids (red), protein (green), and DNA (blue). Scale bars are \qty{20}{\micro\meter}.}
    \label{fig:hyper}
\end{figure}

Retinal pigment epithelial (RPE-1) mammalian cells were imaged sequentially at Raman shifts of 2850, 2934, and \qty{2967}{\per\centi\meter} to demonstrate cavity-enhanced hyperspectral imaging (Fig. \ref{fig:hyper}a-c). These Raman shifts are indicative of lipids, protein, and DNA, respectively \cite{Lu2015}. The datasets are combined into a single hyperspectral image (Fig. \ref{fig:hyper}d). The cell nuclei contain maximal DNA signal (Fig. \ref{fig:hyper}c) and can be identified by the bright blue regions in each cell. The bottom left cell shows two bright lipid structures, a horizontal line near the top and a large droplet near the bottom (Fig. \ref{fig:hyper}a), which show minimal DNA signal. Protein is broadly distributed. The narrow cell projections are visible in all three channels but show weak overall signal and comparable loss to the cell bodies (Extended Data Fig. \ref{fig:phase}). In reality, these three channels are not totally independent due to the overlapping Raman spectra of these biomolecules, and relative or absolute concentrations could be extracted with further calibrations \cite{Oh2022}.

The SNR enhancement is estimated using the same procedure as for Fig. \ref{fig:imaging} to be \qty{3.6(7)}{\decibel} and \qty{4.0(2)}{\decibel} in the absence of electronic noise. A \qty{1.01}{\mega\hertz} amplitude modulation frequency was used to overcome technical noise at lower frequencies, resulting in a \qty{1.2}{\decibel} attenuation of the cavity-enhanced SNR (Fig. \ref{fig:setup}c). These enhanced sensitivities to spatial variations in vibrational spectra and biomolecular concentrations could be used to better differentiate cell types or organelles. 

Cavity-enhanced SRS microscopy provides SNR enhancements of up to \qty{8.3(7)}{\decibel} in spectroscopy and \qty{8.6(1)}{\decibel} in label-free imaging relative to equivalent conventional shot-noise-limited SRS measurements. This technique could also be used to image at higher speeds or with lower doses while maintaining the ESM SNR. In future work, electronic and technical noise sources can likely be further suppressed, and round-trip losses could be reduced. Absolute sensitivity enhancements for all measurement strategies are possible with reduced losses and smaller focal spot sizes. With a reduction in the round-trip loss to 5\% and $T_2 = 0.05$, the total SNR gain could be maximally increased to \qty{12.7}{\decibel} over an ESM. 

Cavity-enhanced measurements can also provide signal enhancements that increase the SNR of measurements that are dominated by technical noise. In contrast, measurements using squeezed states typically require technical noise to be lower than the squeezed noise, which can be technically challenging. Cavity-enhanced measurements are also known to be quantum optimal per input photon---as opposed to per dose---which is relevant when the number of probe particles rather than the number of probe-sample interactions is limited \cite{Belsley2022}. The expected contrast and SNR enhancements per input photon and in the presence of technical noise are quantified in Supplementary Material C.

These metrological gains also apply to measurements of the linear sample absorption at the Stokes wavelength, which is another label-free imaging modality. This information is encoded in the cavity transmission, which is recorded simultaneously with the SRS signal and the single-pass pump transmission (Fig. \ref{fig:setup}a). The dispersive phase shift of the SRS interaction presents an alternative detection scheme \cite{Valensise2019, Qi2024} and can be conveniently measured as a modulation at $f_\text{mod}$ in the error signal that is used to stabilize the cavity length. The information in the phase response regarding the sampled molecular concentration adds to the information obtained from the described intensity measurements \cite{Birchall2020}. The use of this information would further increase the microscope sensitivity with no additional cost in dose. Examples of these measurement modalities are shown in Extended Data Fig. \ref{fig:phase}. 

Cavity-enhanced microscopy is also compatible with other label-free and nonlinear imaging techniques such as stimulated emission, photothermal, transient absorption, and stimulated Brillouin scattering microscopy---some of which have been demonstrated with squeezed light \cite{TriginerGarces2020, Li2022}. We have demonstrated that classical light is sufficient for reaching quantum limits on microscope sensitivity for measurements limited by photodamage. These enhancements could enable measurements of transient biological dynamics or the detection of low-concentration biomolecules in organisms that would otherwise be killed by intense laser light used in nonlinear microscopy. 

\section*{Methods}

\subsection*{Information per dose}
The Fisher information per dose on the stimulated Raman signal $\chi$ for squeezed state probes $\tilde{\mathcal{F}}_\text{SQ} = \eta/[\eta\chi^2 e^{-2r} + (1-\eta\chi)\chi]$, where $\eta$ is the round-trip transmission and $r$ is the squeezing parameter. The maximum Fisher information per dose is achievable with Fock state probes ($r \rightarrow \infty$, quantum limit), $\tilde{\mathcal{F}}_Q = \eta/[\chi(1-\chi\eta)]$ \cite{Allen2020}. This is also the Fisher information for cavities with $T_1 \rightarrow 0$ and $T_2 \rightarrow 1-\eta\chi$. For coherent state probes in single-pass ($r = 0$), $\tilde{\mathcal{F}}_\text{SP} = \eta/\chi$. For suboptimal $T_1$ and $T_2$, the cavity-enhanced information $\tilde{\mathcal{F}}_\text{CE} = G^2\tilde{\mathcal{F}}_\text{SP}$ (Eqn. \ref{gain}). The numerical results in Fig. \ref{fig:setup}b were calculated for $\chi = 1 + 10^{-4}$, which is typical for SRS \cite{Audier2020}. Further details are given in Supplementary Material A. 

\subsection*{Equivalent single-pass measurements}

An ideal ESM would be conducted with the same optical dose under identical alignment conditions with the same signal detection and processing chain. For $T_2 = 0.104$, measurements performed at equivalent dose would require a photodiode showing a linear response over an order-of-magnitude of Stokes powers, which is challenging for measurements of short pulses. Furthermore, removing or circumventing the cavity-optics for high-dose single-pass measurements would alter the Stokes alignment and, critically, the spatial and temporal overlap of the pump and Stokes beams at the optical focus. 

For a fair comparison with single-pass measurements, the constant-dose condition is abandoned to maintain the optical alignment and the detection conditions between single-pass and cavity-enhanced measurements. Single-pass measurements are conducted by rerouting the Stokes beam to the detector $D_S$ after the second microscope objective by inserting two additional mirrors. This bypasses one cavity concave mirror and $M_2$ but otherwise uses the same optics and detector. The single-pass dose is measured after these mirrors so that any additional optical losses are included in the scaling of the signal to the constant-dose condition. The alignment of the single-pass beam onto the photodiode is verified by ensuring that the measured optical power produces the expected DC voltage for the photodiode responsivity and gain. 

The measured single-pass signal is scaled by the ratio of the cavity-enhanced dose to the single-pass dose to estimate the single-pass signal that would be observed in an ESM. The measured power of the shot noise is scaled by the ratio of the cavity-enhanced optical output power to the cavity-enhanced dose to estimate the single-pass shot noise that would be observed for the same dose. This assumes that the equivalent single-pass measurement could be conducted with negligible technical noise, which provides the most conservative estimate of the SNR enhancement due to the optical cavity. The cavity-enhanced measurement SNR is then compared to the SNR that would be observed under identical measurement conditions for single-pass measurements using the same optical dose. 

\subsection*{Cavity-enhanced SRS microscope}
The pump and Stokes beams are generated by a dual-output femtosecond laser with $f_\text{rep} \sim $ \qty{80}{\mega\hertz} (Spectra-Physics InSight X3). The Stokes pulses have a center wavelength of \qty{1045}{\nano\meter} and an initial duration of about \qty{200}{\femto\second}; the pump pulses have a tunable center wavelength and an initial duration of about \qty{120}{\femto\second}. The power in each beam is independently controlled with a half-wave plate and polarizer. 

The Stokes pulses are chirped by \qty{-0.133}{\square\pico\second} with a transmission grating pair (Coherent T-1000-1040) before being phase-modulated at \qty{11.7}{\mega\hertz} with a resonant EOM (New Focus Model 4001). The pulses are sent through a \qty{0.2}{\meter} single-mode, polarization-maintaining fiber (PM-980) to improve the spatial mode profile. After the fiber, three lenses are used to match the Stokes beam mode to the TEM$_{00}$ mode of the optical cavity.

The ring cavity length is stabilized such that $f_\text{FSR} = f_\text{rep}$ using the Pound-Drever-Hall technique \cite{Drever1983}. The reflected light at $M_1$ is monitored with a balanced photodiode (Thorlabs DET36A2). Unlike in other experiments that stabilize mode-locked lasers to optical cavities \cite{Krueger1995, Jones2001}, every longitudinal laser mode is detected on a single detector. A low-pass filter isolates the beat note between the phase-modulation sidebands and the carrier, and this signal is amplified by approximately \qty{50}{\decibel} (Mini-Circuits ZFL-500LN+, ZFL-1000LN+) and demodulated with an \qty{11.7}{\mega\hertz} local oscillator with a double-balanced mixer (Mini-Circuits ZFM-2-S+). The resulting error signal is used to feedback to the position of a piezo-mounted mirror using a high-speed servo controller (New Focus LB1005-S) and high-voltage amplifier (PiezoDrive PX200) to stabilize the cavity length. 

The Stokes beam is focused to a waist with a $1/e^2$ intensity radius of \qty{1.4(2)}{\micro\meter} at the sample plane between two 0.6 NA microscope objectives (Thorlabs LMH-50X-1064). The Stokes beam 1/$e^2$ intensity radius at the back aperture of the microscope objectives is \qty{1.05}{\milli\meter}. The concave mirrors have radii of curvature of \qty{3000}{\milli\meter}. For near-quantum-optimal performance, we choose $T_1 = 0.004$ and $T_2 = 0.104$. For a finesse of 30, the round-trip loss $1 - \eta = 0.09$, constraining the average loss per microscope objective to $<5\%$.  After the sample and objectives, a portion of the circulating Stokes pulse is transmitted through $M_2$ and detected. 

The cavity is also characterized by the cooperativity $C$, also called the Purcell factor, which quantifies the fraction of dipole emission from a scatterer into the cavity mode relative to the total emission into all space, and $C > 1$ demarcates the strong-coupling regime of cavity quantum electrodynamics. For a traveling wave cavity, $C = 6F/(\pi k^2 w^2)$ where $k$ is the light wavevector and $w$ is the mode waist \cite{TanjiSuzuki2011}. For a waist of \qty{1.4(2)}{\micro\meter} in PBS ($n = 1.33$), $k = 2\pi n/\lambda$, and $F = 30$, $C \approx 0.5$. At room temperature, the short coherence times of vibrational modes limit the effective cooperativity \cite{Huemmer2016}, making cooperative effects negligible in our experiments. 

The pump pulses are also chirped by \qty{-0.133}{\square\pico\second} with a transmission grating pair (Coherent T-1400-800). A translation stage is used to temporally overlap the Stokes and pump pulses and to adjust the measured Raman shift via spectral focusing. For the applied dispersion, a \qty{1}{\pico\second} increase in the pump delay (\qty{300}{\micro\meter} increase in path length) corresponds to a shift of approximately \qty{-40}{\per\centi\meter}. The spectral resolution is \qty{19}{\per\centi\meter}, defined as the FWHM of the \qty{2913}{\per\centi\meter} peak of DMSO (Extended Data Fig. \ref{fig:spontaneous}). The conversion from delay stage position to Raman shift is calibrated using the known Raman shifts of the C-H stretching modes of common organic solvents and is referenced to DMSO spectra measured on the same day as the imaging experiments. The round-trip dispersion of the cavity has a negligible effect on the measurements due to the large initial chirp, low finesse, and moderate Stokes spectral bandwidth.

The pump beam is amplitude-modulated at $f_\text{mod}$ by an AOM (Isomet 1206-C), and the first-order diffracted beam is used to achieve a modulation depth of approximately 100\%. A variable beam expander, a \qty{400}{\milli\meter} lens, and two mirrors are used to control the size and three-dimensional position of the pump beam waist on the sample. The pump beam is reflected off of the intracavity dichroic mirror onto the cavity mode axis at the back aperture of the first objective. The dichroic also serves as a polarizer for the Stokes beam and defines the polarization of the circulating Stokes pulse. Both beams are linearly polarized with parallel polarizations. 

All optical powers are measured using a power meter (Thorlabs S121C) with a measured linear response up to at least \qty{80}{\milli\watt}. For SRS measurements, the light transmitted through $M_2$ is long-pass filtered (Thorlabs FELH1000) to remove any residual pump light and detected on an amplified photodiode (Thorlabs PDA100A2 for DMSO, HeLa cell images; PDA20X2 for hyperspectral images). For cavity-enhanced measurements, the average optical power at the detector is $1/T_2$ lower than the average Stokes dose. Approximately 1\% of the light leaving the cavity is picked off and sent to a biased photodiode (Thorlabs DET20C2) to monitor the cavity transmission. The pump light is transmitted by the first concave mirror after the objectives and is monitored with another biased photodiode (Thorlabs DET36A) to measure the sample absorption at the pump wavelength.

All signals are synchronously digitized by a high-speed lock-in amplifier (Zurich UHFLI-600). The SRS signal at $f_\text{mod}$ is digitally demodulated and filtered with a fourth-order low-pass filter. The sampling rate for all SRS measurements was 13.7 kSa/s. The Stokes and pump transmissions were sampled at 400 kSa/s.

\subsection*{Sample preparation}
Samples are sealed between two 1.5H glass coverslips with a \qty{120}{\micro\meter}-thick adhesive spacer (Grace Bio-Labs SecureSeal). The outside faces of the coverslips have anti-reflection coatings with R $<$ 0.2\% for 750 to \qty{1050}{\nano\meter} (ios Optics) to minimize losses at both the pump and Stokes wavelengths. The inside faces are uncoated. 

RPE-1 (CRL-4000, ATCC) and HeLa (CRM-CCL-2, ATCC) cells were cultured and maintained at 37 °C, and 5\% CO2 environment using Dulbecco’s Modified Eagle Medium (DMEM) (Gibco, Inc) supplemented with 10\% (v/v) Fetal Bovine Serum (FBS) (Life Technologies), and 1\% (v/v) penicillin-streptomycin (Gibco). For imaging, the cells were grown overnight on the uncoated coverslip surface. 

Prior to imaging, coverslips and adherent cells were removed from the growth medium and gently rinsed three times with 1X phosphate-buffered saline (PBS) (Gibco). The chamber formed by the adhesive spacer and coverslip was then filled with PBS prior to sealing with the second coverslip. All cell images were acquired under ambient conditions at room temperature. 

\subsection*{Noise measurements}
Noise power spectral densities are calculated from the measured lock-in output using Welch's method (Extended Data Fig. \ref{fig:noise}b). Electronic noise is measured with the Stokes beam blocked. Total noise is measured with the Stokes and pump beams not temporally overlapped. The mean power spectral density is calculated between \qty{30}{\hertz} and $f_\text{NEP}/2$, where $f_\text{NEP}$ is the noise-equivalent-power bandwidth of the measurement (0.078/$t_c$, where $t_c$ is the lock-in time constant). The shot noise is the difference between the total noise power and the electronic noise power. Reported noises are calculated by multiplying the densities by the signal measurement $f_\text{NEP}$ and reported with the number of time-domain samples $n$ used to calculate the power spectral density and the number of frequencies $m$ used to compute the average power.

\subsection*{Spectroscopy experiments and analysis}
The pump beam center wavelength was \qty{795}{\nano\meter}, and the average pump dose was \qty{14}{\milli\watt}. The lock-in time constant was \qty{100}{\micro\second}. The pump delay stage was driven at a velocity of \qty{150}{\micro\meter\per\second}, corresponding to a spectral scan rate of \qty{-40}{\per\centi\meter\per\second} due to spectral focusing. This scan speed is slow compared to both the cavity lifetime (\qty{60}{\nano\second}) and the lock-in amplifier settling time (\qty{1}{\milli\second}). In spectral focusing, the measured spectra differ from the spontaneous Raman spectrum due to the shift-dependent overlap of the pump and Stokes pulse intensity envelopes (Extended Data Fig. \ref{fig:spontaneous}).

The SRS signals are isolated and measured in a single quadrature that is in-phase with the amplitude modulation at $f_\text{mod} = $ \qty{210}{\kilo\hertz}. The Raman signal is calculated from each sequentially measured spectrum as the mean lock-in output voltage over a shift in the stage position of \qty{2.2}{\micro\meter} around the peak at \qty{2913}{\per\centi\meter} ($n = 15$ samples). The zero-dose measurement is taken with the Stokes beam blocked. The average cavity-enhanced signals are fit with an unweighted linear regression ($n = 12$ doses), yielding a slope \qty{7.9(3)}{\milli\volt\per\milli\watt} dose and intercept $4(3)\times 10^{-2}$ \unit{\milli\volt}. Uncertainties are 95\% confidence intervals. The mean signal enhancement, the mean SNR enhancement, and the standard errors are given in the text ($n = 11$ doses). 

The noise powers measured at each dose (\qty{0}{\milli\watt}, $n = 24,000$, $m = 98$; $>$\qty{0}{\milli\watt}, $n = 24,000$, $m = 214$) are fit with an unweighted linear regression ($n = 12$ doses) to estimate the dose-dependent noise. The total noise power is the sum of the electronic noise \qty{0.16(4)}{\square\micro\volt} and the shot noise for the detector $D_S$ illuminated by the fraction $T_2$ of the Stokes dose \qty{0.27(4)}{\square\micro\volt\per\milli\watt}. Uncertainties are 95\% confidence intervals. For shot noise, the theoretical power spectral density is $2A^2 eR P_S$, where $e$ is the elementary charge, $A$ is the electronic gain, $R$ is the photodiode responsivity, and $P_S$ is the average Stokes power at $D_S$. The photodiode gain $A\sqrt{R} =$ \qty{1.06(2)}{\kilo\volt \per\watt\tothe{0.5}\per\ampere\tothe{0.5}} for this measurement. This is less than the photodiode specification $A\sqrt{R} \approx$ \qty{1.73(3)}{\kilo\volt \per\watt\tothe{0.5}\per\ampere\tothe{0.5}}, where the uncertainty is the manufacturer's specified gain variation. The difference is due to additional electronic losses in the detection chain, the rapidly changing specification for $R$ around \qty{1045}{\nano\meter}, and an uncertain absolute quantum efficiency for the power meter, all of which are common-mode to both single-pass and cavity-enhanced measurements. 

Measurement noise can also be characterized in the frequency domain. The noise amplitude spectral density is measured as a function of the Stokes power and lock-in demodulation frequency. The average noise power spectral density in the in-phase quadrature from \qty{204.8}{\kilo\hertz} to \qty{216.0}{\kilo\hertz} ($n = 9$ frequencies) is estimated for the same optical powers and fit with an unweighted linear regression ($n = 13$ doses) (Extended Data Fig. \ref{fig:noise}) to estimate the electronic noise floor \qty{0.21(6)}{\square\micro\volt} and shot noise \qty{0.30(5)}{\square\micro\volt\per\milli\watt}. The total noise power is again calculated by scaling the power spectral density by $f_\text{NEP}$. Noise characterization using the lock-in output time series is preferred to reduce systematic uncertainties that result from averaging over noise powers that are measured at frequencies outside of the measurement filter bandwidth. 

\subsection*{Cavity-enhanced imaging experiments and analysis}
The pump center wavelength was \qty{798}{\nano\meter}, and the average pump dose was \qty{60}{\milli\watt}. The average single-pass Stokes dose is \qty{138}{\micro\watt}, and the average cavity-enhanced Stokes dose is \qty{8}{\milli\watt}. The lock-in time constant was \qty{250}{\micro\second}. 

The sample was moved in \qty{1}{\micro\meter} steps in the vertical direction, and SRS was measured as the sample was scanned through the cavity waist at \qty{100}{\micro\meter\per\second} in the horizontal direction, giving a dwell time in a \qtyproduct{1 x 1}{\micro\meter} pixel of \qty{10}{\milli\second}. Each image was acquired in 5 minutes, and the time delay between image starts was 15 minutes. The Raman signals are isolated and measured in a single quadrature that is in-phase with the amplitude modulation at $f_\text{mod} =$ \qty{210}{\kilo\hertz}. 

Prior to enhancement estimation, a baseline correction is applied to each row by subtracting a linear fit to the pixels with signals below a global voltage threshold. The images are linearly interpolated onto a coarser pixel grid to lower the measurement sensitivity to technical noise and potential small sample drifts between images. The signal enhancement is estimated as the reciprocal of the slope that is fit with a Deming regression to the scatter plot of the measured single-pass versus cavity-enhanced signals for each pixel, where the error variances in the signals are estimated from nonlinear least squares Gaussian fits to the Raman signals near zero, which represent background pixels. The quoted enhancement is 48.3(4) for an interpolated pixel size of \qty{2.25}{\square\micro\meter}. The uncertainty is the standard deviation in the enhancement calculated from 200 regressions that are run with the baseline-correction signal thresholds and the measurement variance ratios allowed to independently vary randomly by up to 30\%. For interpolated pixel sizes of 1, 2.25, and \qty{4}{\square\micro\meter} ($n = 22,154$; 9,940; and 5,565 pixels), the enhancement is 48.2(4), 48.3(4), and 48.9(5), respectively. The standard error on each individual fitted enhancement is estimated using the jackknife method \cite{Linnet1990} to be 0.6, 0.9, and 1, respectively. 

The electronic noise is \qty{0.02(1)}{\square\micro\volt} ($n = 20,000$, $m=76$). The shot noise \qty{.16(2)}{\square\micro\volt} is calculated from a measurement of the total noise ($n = 20,000$, $m = 30$). Uncertainties are standard errors on the means. An ESM at constant dose would have a shot noise of \qty{1.6(2)}{\micro\volt\tothe{2}}. The total SNR enhancement is calculated by dividing the estimated signal enhancement by the ratio of the cavity-enhanced dose to the single-pass dose and by the square root of the ratio of the powers of the cavity-enhanced noise and the single-pass shot noise. 

\subsection*{Hyperspectral imaging experiments and analysis}
The pump beam center wavelength was \qty{798}{\nano\meter}, and the average pump dose was \qty{46}{\milli\watt}. The average single-pass Stokes dose was \qty{160}{\micro\watt}, and the average cavity-enhanced Stokes dose was \qty{12}{\milli\watt}. The lock-in time constant was 500 (980) \unit{\micro\second} for the cavity-enhanced (single-pass) measurements. A longer time constant was used to increase the single-pass SNR for quantitative comparisons. 

The sample was moved in \qty{1}{\micro\meter} steps in the vertical direction, and SRS was measured as the sample was scanned through the waist at \qty{100}{\micro\meter\per\second} in the horizontal direction, giving a dwell time in a \qtyproduct{1 x 1}{\micro\meter} pixel of \qty{10}{\milli\second}. Each \qtyproduct{125 x 135}{\micro\meter} image was acquired in 4 minutes, and the time delay between each image start was 9 minutes. The cavity-enhanced signal was isolated and measured in a single quadrature that is in-phase with the amplitude modulation at $f_\text{mod} = $ \qty{1.01}{\mega\hertz}. The single-pass signal was not perfectly in-phase, so the signal is measured as the magnitude of the two baseline-corrected quadrature signals. 

For each displayed image, a baseline correction is applied to each row by subtracting a linear fit to the pixels with signal values below a global voltage threshold. To generate the hyperspectral image, the signals in each background-subtracted image are normalized and then used as the red, green, or blue component of the RGB color image. 

For the SNR enhancement calculations, the analysis is restricted to a \qtyproduct{70 x 90}{\micro\meter} region containing the lower-left cell. The other two cells were photodamaged after hyperspectral imaging and prior to single-pass imaging. Cavity-enhanced and single-pass images recorded at \qty{2934}{\per\centi\meter} (collected 1.5 hours apart) are analyzed in the same way as for Fig. \ref{fig:imaging}. The quoted enhancement is 37(1) for an interpolated pixel size of \qty{2.25}{\square\micro\meter}. The uncertainty is the standard deviation in the enhancement calculated from 200 regressions that are run with the baseline-correction signal thresholds and the measurement variance ratios allowed to independently vary randomly by up to 30\%. For interpolated pixel sizes of 1, 2.25 and \qty{4}{\square\micro\meter}, the enhancement is 40(1), 37(1), and 37(1), respectively. The standard error on each individual fitted enhancement ($n = 7,500$; 4,050; and 1,922 pixels, respectively) is estimated using the jackknife method to be 1, 1, and 2, respectively. 

The electronic noise is \qty{0.011(7)}{\square\micro\volt} ($n = 17,162$, $m = 75$). The shot noise \qty{0.12(1)}{\square\micro\volt} is calculated from a measurement of the total noise ($n = 20,000$, $m = 41$). Uncertainties are standard errors on the means. An ESM at constant dose would have a shot noise of \qty{1.2(1)}{\square\micro\volt}. The total SNR enhancement is calculated by dividing the estimated signal enhancement by the ratio of the cavity-enhanced dose to the single-pass dose and by the square root of the ratio of powers of the cavity-enhanced noise and the estimated single-pass shot noise. The single-pass shot noise is restricted to that which would be measured in a single quadrature, which is a lower bound on the single-pass measurement noise. 

\subsection*{Estimation of $\eta$}
The finesse $F$ is measured by fitting Lorentzian functions to the cavity transmission peaks, recorded as the cavity length is scanned (Extended Data Fig. \ref{fig:waist}), using nonlinear least squares ($n = 10,000$ positions per peak). Reported $F$ values are calculated from three separately measured pairs of transmission peaks, and the mean and standard error are reported. The finesse depends on the round-trip loss $1-\eta$ for known $T_1$ and $T_2$,
\begin{equation}
    F = \frac{\pi}{2 \ \text{arcsin}\left(\frac{1-\sqrt{\rho}}{2\rho^{1/4}}\right)},
\end{equation}
where $\rho = (1-T_1)(1-T_2)\eta$. Once the loss at one pixel is known, the cavity intensity transmission through $M_2$, 
\begin{equation}
    T = \frac{T_1 T_2 \eta \chi}{\left(1-\sqrt{(1-T_1)(1-T_2)\eta\chi} \right)^2},
\end{equation}
is evaluated numerically to calculate the transmission for all pixels (Extended Data Fig. \ref{fig:phase}c). 

\subsection*{Cavity waist characterization}
Two lineouts ($n = 2,768$ pixels, each) from the cavity-enhanced transmission image of RPE-1 cells (Extended Data Fig. \ref{fig:phase}c), are fit with the sum of four Gaussians and a linearly changing baseline (Extended Data Fig. \ref{fig:waist}c) using nonlinear least squares. The waist is estimated as the intensity half width at $1/e^2$ from the fitted Gaussians satisfying the condition that the intensity half width at $1/e^2 < $ \qty{2}{\micro\meter} ($n = 4$ waists) to be \qty{1.4(2)}{\micro\meter}. The uncertainty is the standard error on the mean. 

\section*{Supplementary Materials}
\subsection*{A Fisher information calculations}

The Fisher information $\mathcal{F}$ is a statistical measure of the minimum variance that can be achieved by an unbiased estimator of an unknown parameter characterizing the distribution of a set of observations. In optical microscopy, intensity measurements are used to estimate sample properties such as refractive index, absorption coefficient, or molecular concentration. The sample alters the quantum state of the light, and each measurement and estimation strategy for a given property can be quantified by $\mathcal{F}$. The quantum Fisher information $\mathcal{I}$ quantifies the information available in the quantum state, independent of the measurement strategy, and thus bounds $\mathcal{F}$ from above according to the quantum Cramer-Rao bound,
\begin{equation}
    \mathcal{I}(\chi) \geq \mathcal{F}(\chi) \geq \frac{1}{\nu \ \text{Var}(\chi)},
\end{equation}
where Var($\chi$) is the variance in the estimate of the unknown parameter $\chi$, and $\nu$ is the number of times that identical experiments are repeated \cite{Braunstein1994}. 

In SRS microscopy, the stimulated Raman signal $\chi$ is weak compared to the measurement loss $1-\eta$, resulting in a net loss of light $\eta \chi < 1$. Under these conditions, SRS microscopy reduces to a loss estimation problem. The Raman signal and the loss are defined by their action on a coherent state, $\ket{\alpha}$ $\rightarrow$ $\sqrt{\eta \chi }\ket{\alpha}$, where the optical dose $|\alpha|^2$ is the mean number of photons in the state. When the total dose is limited by the onset of sample damage, a quantum-optimal measurement maximizes the Fisher information per optical dose $\tilde{\mathcal{F}} = \tilde{\mathcal{I}}$. It is assumed that there is no dose cost to measuring $\eta$ such that it can be known to arbitrarily high precision relative to $\chi$. 

For single-pass measurement strategies, 
\begin{equation}
    \tilde{\mathcal{F}}(\chi, \eta, \sigma_\psi) = \frac{\eta}{\eta \chi^2  \sigma_\psi + (1-\eta \chi)\chi},
\end{equation}
where $\sigma_\psi = $ Var($N_0$)/$N_0$ is the Fano factor of a general input quantum state with $N_0$ photons on average \cite{Allen2020}. For coherent states, $\sigma_\psi$ = 1, and $\tilde{\mathcal{F}_C} = \eta/\chi $. For Fock states $\sigma_\psi = 0$, and $\tilde{\mathcal{F}_Q} = \tilde{\mathcal{I}} =  \eta/[\chi(1-\eta\chi)]$.  

For a squeezed state, $\sigma_\psi$ depends on both the absolute displacement and the squeezing parameter $r$. The squeezing in \unit{\decibel} is $20 r/\text{ln}(10)$. The average number of photons $N_0 = |\alpha|^2 + \text{sinh}^2r$. When the squeezing is small, which is typical for the SRS experiments under consideration, $|\alpha|^2 \gg \text{sinh}^2 r$. In this limit, $\sigma_\psi  = e^{-2r}$ for all displacements. The maximum information enhancement for non-classical probes relative to coherent state probes, 
\begin{equation}
    \frac{\tilde{\mathcal{F}}_Q (\chi,\eta)}{\tilde{\mathcal{F}}_C(\eta,\chi)} = \frac{1}{1-\eta\chi}, 
    \label{qLimit}
\end{equation}
can be achieved with Fock state probes in a single-pass configuration. For Fock states, the average dose is the total dose. This is no longer true for squeezed or classical states of light, for which only average doses can be specified. Further quantum advantages might be available for absolute, rather than average, doses. 

For an optical ring cavity on resonance \cite{Yariv2000}, the average number of photons $N$ transmitted through $M_2$ is 
\begin{equation}
    {N} = \frac{T_1 T_2 \eta \chi N_0}{\left(1-\sqrt{(1-T_1)(1-T_2)\eta \chi}\right)^2}, 
\end{equation}
and the average dose $D$ is 
\begin{equation}
    D = \frac{T_1 N_0}{\left( 1- \sqrt{(1-T_1)(1-T_2)\eta \chi}\right)^2}, 
\end{equation}
assuming that losses occur after the probe--sample interaction. An estimate of $\chi$ is limited by the variance in $N$, which results from shot noise, 
\begin{equation}
    \text{Var}(\chi) = \frac{\text{Var}(N)}{\nu \ (dN/d\chi)^2 } = \frac{N}{\nu \ (dN/d\chi)^2}. 
\end{equation}

The Fisher information per dose, 
\begin{equation}
    \tilde{\mathcal{F}}_\text{CE} = \frac{T_2 \eta}{\chi\left( 1 - \sqrt{(1-T_1)(1-T_2)\eta \chi} \right)^2},
\end{equation}
is optimized for a highly reflective input coupler $T_1 \rightarrow 1$, and $T_2 \rightarrow 1-\eta\chi$ for cavity impedance matching. In this limit, the information enhancement saturates the quantum limit (Eqn. \ref{qLimit}), 
\begin{equation}
    \frac{\tilde{\mathcal{F}}_\text{CE}}{\tilde{\mathcal{F}}_C} = \frac{T_2}{\left(1-\sqrt{(1-T_1)(1-T_2)\eta \chi} \right)^2}\rightarrow  \frac{1}{1-\eta\chi}. 
\end{equation}

This analysis only includes sample and microscope objective losses. Finite detection efficiencies are considered in Supplementary Material B. We have also assumed that the undetected beam dose does not provide an additional constraint on the measurement. As all measurements benefit from increased intensity in the undetected pump beam in stimulated Raman gain, the average pump dose should be maximally increased and is assumed to be fixed across all measurement strategies. 

The optimal SNR for classical single-pass measurements constrained by a total dose $N_P + N_S$ is achieved in stimulated Raman gain for $N_P = 2N_S$ \cite{Moester2015, Audier2020}. For a cavity-enhanced measurement, this condition still holds, but the average detected number of Stokes photons is $T_2 \eta N_S$. The amplitude of the photon shot noise is reduced by $\sqrt{T_2}$. 

\subsection*{B Quantum limits on sensitivity}
In addition to sample and microscope objective losses, imperfect detection degrades measurement performance. For a detection quantum efficiency $\eta_d$, the ratio of the cavity-enhanced SNR to the ESM SNR is unchanged and saturates the quantum limit on information enhancement $1/(1-\eta\chi)$ that could be achieved with $\eta_d = 1$ because both SNRs are proportional to $\sqrt{\eta_d}$. Quantum probes, by contrast, are limited in their information enhancement to a lower value $1/(1-\eta\eta_d\chi)$. However, the absolute sensitivity, rather than the sensitivity enhancement, of cavity-enhanced measurements with finite detector quantum efficiency is reduced below the absolute quantum limit that could be achieved with quantum states and perfect detection. For $\eta_d \sim 0.5$, which is typical for the photodiodes used in SRS experiments, the absolute cavity-enhanced SNR under otherwise equivalent measurement conditions is reduced by \qty{3}{\decibel} relative to a perfect quantum measurement. Additional technical noise sources also limit classical and quantum measurement sensitivities relative to absolute quantum limits, requiring mitigation strategies such as signal modulation and lock-in detection. 

\subsection*{C Alternative measurement constraints}
For a single-pass measurement using $N$ input Stokes photons to detect a small loss (transmission, stimulated Raman loss) or gain (stimulated Raman gain) $\chi = 1 + \delta\chi$ in the presence of a loss $1-\eta$, the signal $S_\text{SP}$, shot noise $\sigma_\text{SP}$, and SNR$_\text{SP}$ are
\begin{align}
    S_\text{SP} & = \eta N \delta \chi, \\
    \sigma_\text{SP} & = \sqrt{\eta N}, \\
    \text{SNR}_\text{SP} & = \delta\chi \sqrt{\eta N}, 
\end{align}
under the simplifying assumption that $\chi$ has a negligible effect on the shot noise. 

For a measurement enhanced by a resonant cavity, the dose $D$ is altered, 
\begin{equation}
    D = \frac{T_1 N}{\left(1-\sqrt{\eta(1-T_1)(1-T_2)}\right)^2},
\end{equation}
along with the signal, noise, and SNR, 
\begin{align}
    S & = \frac{T_1 T_2 \eta N \delta \chi}{\left(1-\sqrt{\eta(1-T_1)(1-T_2)}\right)^3}, \\
    & = \frac{D T_2 \eta \delta \chi}{1-\sqrt{\eta(1-T_1)(1-T_2)}}\\
    \sigma & =  \sqrt{DT_2\eta}, \\
    \text{SNR} & = \frac{\delta \chi\sqrt{D T_2 \eta}}{1-\sqrt{\eta(1-T_1)(1-T_2)}}. 
\end{align}
where $T_1$ ($T_2$) is the cavity input (output) coupler intensity transmission and $\chi$ is assumed to also have a negligible effect on the dose. 

The main text presented a dose-limited measurement at the shot-noise limit. The theoretical gain in SNR relative to an equivalent single-pass measurement
\begin{equation}
    G = \frac{\sqrt{T_2}}{1-\sqrt{\eta(1-T_1)(1-T_2)}}
\end{equation}
is 3.26 (\qty{10.3}{\decibel}) for the cavity with $T_1 = 0.004$, $T_2 = 0.104$, and $\eta = 0.91$. At the quantum limit ($T_1 = 0$,  $T_2 = 1-\eta = 0.09$), the gain is 3.33 (\qty{10.5}{\decibel}). 

For measurements limited by a constant technical noise floor, the SNR gain reduces to the signal gain. At constant dose, 
\begin{equation}
    \frac{S}{S_\text{SP}} = \frac{T_2}{1-\sqrt{\eta(1-T_1)(1-T_2})},
\end{equation}
which is 1.05 (\qty{0.4}{\decibel}) for $\eta =0.91$. This regime applies to the measurements of the cavity transmission. The enhancement is maximized for $T_1 \rightarrow 0$ and $T_2 \rightarrow 2(1-\eta^{-1}+\eta^{-1} \sqrt{1-\eta})$, at which point the gain is 1.54 (\qty{3.7}{\decibel}). For quantum states of light, there would be no increase in SNR. 

This enhancement is the result of an increase in the signal contrast, defined as the ratio of the number of signal photons to the total number of photons at the detector. The cavity enhanced contrast $S/DT_2 \eta$ is enhanced over the single-pass contrast by $1/\left(1-\sqrt{\eta(1-T_1)(1-T_2)}\right)$, which is 10.1 (\qty{20.1}{\decibel}) for $\eta = 0.91$. For $\eta = 0.85$, the expected contrast enhancement is 7.7 (\qty{17.8}{\decibel}) (Extended Data Fig. \ref{fig:phase}b-c). 

The measurement SNR is also increased when the number of input photons, rather than the total dose, is limited. For a shot-noise-limited measurement, 
\begin{equation}
    G = \frac{\sqrt{T_1T_2}}{\left( 1- \sqrt{\eta(1-T_1)(1-T_2)} \right)^2},
\end{equation}
which is 2.09 (\qty{6.4}{\decibel}) for the cavity with $\eta = 0.91$. This expression is optimized when $|T_1| = |T_2| = \eta^{-1/2}-1$, for which the gain is 5.69 (\qty{15.1}{\decibel}). 

For measurements limited by the number of input photons and technical noise, there is no SNR increase for quantum states of light, but the cavity alters the signal, 
\begin{equation}
    \frac{S}{S_\text{SP}} = \frac{T_1T_2}{\left(1-\sqrt{\eta(1-T_1)(1-T_2)} \right)^3},
\end{equation}
which is 0.43 (\qty{-7.3}{\decibel}) for $\eta = 0.91$. The enhancement is maximized for $T_1 = T_2 = 2(\eta^{-1/2}-1)$, for which the gain is 5.06 (\qty{14.1}{\decibel}). 

\backmatter

\bmhead{Author contributions} S. C. B., J. L. R., and T. Y. developed the microscope. J. L. R. performed the experiments and wrote software to acquire and analyze the data. S. M. prepared cells for imaging and conducted spontaneous Raman scattering experiments. J. L. R., S. C. B., S. M., and M. A. K. analyzed the data. J. L. R., S. C. B., and M. A. K. wrote the manuscript. All authors revised the manuscript. 

\bmhead{Data availability}

Raw datasets supporting this study have been deposited to Zenodo (DOI: 10.5281/zenodo.17794025). Additional data is available from the corresponding author on reasonable request. 

\bmhead{Code availability}

The computer code generated to analyze the datasets in this study is available from the corresponding author on reasonable request. 

\bmhead{Supplementary information}

Supplementary information is available for this paper.  

\bmhead{Acknowledgments}

We thank Stewart Koppell, Brannon Klopfer, Guglielmo Panelli, Erik Porter, Rose Knight, Nahal Bagheri, and Devin Dean for helpful discussions. We thank Callista Yee, Taylar Hammond, and Luke Qi for providing additional test samples. This work was supported by the US Department of Energy, Office of Science, Office of Biological and Environmental Research under award number DE-SC0023076. 

\section*{Declarations}

The authors declare no competing interests.


%\bibliography{bibliography}% common bib file

\begin{appendices}

\newpage

\section*{Extended Data}

\begin{figure}[h]
    \centering
    \makebox[\textwidth]{\includegraphics[width=18cm]{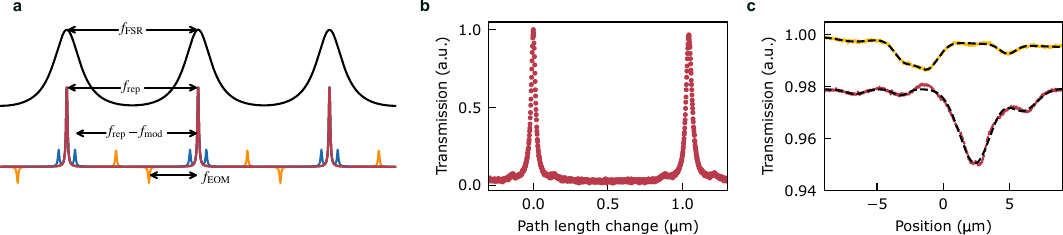}}
    \caption{\textbf{Extended Data Figure 1} Cavity characterization. (a) Schematic of the experiment in the frequency domain. Each mode of the Stokes beam (spacing $f_\text{rep} \sim $ \qty{80}{\mega\hertz}) is on resonance with a longitudinal cavity mode (free spectral range $f_\text{FSR} \sim $ \qty{80}{\mega\hertz}), offset for clarity. Phase-modulation sidebands (yellow) are placed on each laser mode by an EOM to stabilize the cavity length. Stimulated Raman scattering transfers amplitude-modulation sidebands at $f_\text{mod}$ (blue) from the pump beam to the Stokes beam, and these sidebands are attenuated by the transfer function of the cavity. (b) Cavity transmission with a DMSO sample inserted, measured as the cavity length is scanned with the piezoelectric transducer. (c) Gaussian fits (dashed, black) to transmission lineouts of RPE-1 cells (Extended Data Fig. \ref{fig:phase}c). The mean half width at $1/e^2$ for the fitted Gaussians with half widths at $1/e^2 < $ \qty{2}{\micro\meter} is \qty{1.4(2)}{\micro\meter}, where the uncertainty is the standard deviation of the width measurements.}
    \label{fig:waist}
\end{figure}

\newpage

\begin{figure}[h]
    \centering
    \includegraphics[width = 9cm]{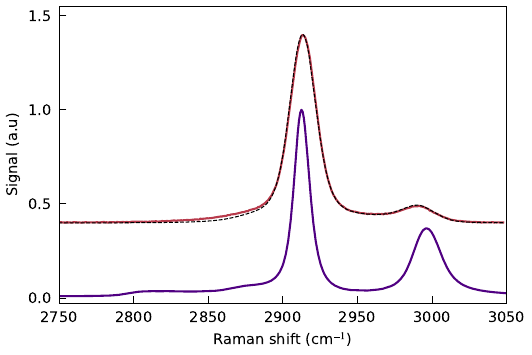}
    \caption{\textbf{Extended Data Figure 2} Comparison with spontaneous Raman scattering spectroscopy. The spontaneous Raman spectrum of DMSO (purple), measured on a commercial microscope (Horiba Labram HR Evolution, \qty{532}{\nano\meter} excitation, 600 lines \unit{\per\milli\meter}), is convolved with a Gaussian filter (scipy gaussian\_filter) with a standard deviation of 6 cm$^{-1}$. The broadened spectrum is then weighted by multiplication with a Gaussian function with a standard deviation of 38 cm$^{-1}$ that is centered at 2923 cm$^{-1}$ (dashed, blue) for comparison with the measured cavity-enhanced spectrum (red) acquired with spectral focusing. Offset for clarity.}
    \label{fig:spontaneous}
\end{figure}

\newpage

\begin{figure}[h]
    \centering
    \makebox[\textwidth]{\includegraphics[width=18cm]{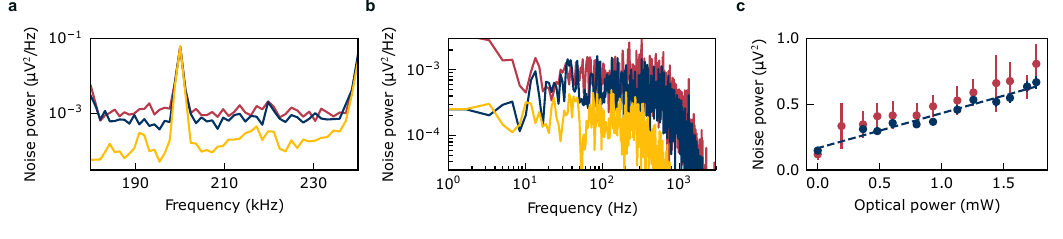}}
    \caption{\textbf{Extended Data Figure 3} Noise characterization for DMSO spectroscopy. (a) Noise power spectral density, measured as a function of lock-in amplifier demodulation frequency, for Stokes powers at $D_S$ of \qty{1.8}{\milli\watt} (red), \qty{1.1}{\milli\watt} (blue), and \qty{0}{\milli\watt} (orange). (b) Noise power spectral density measured from the time-domain output of the lock-in amplifier after demodulation at \qty{210}{\kilo\hertz} for Stokes powers at $D_S$ of \qty{1.8}{\milli\watt} (red), \qty{1.1}{\milli\watt} (blue), and \qty{0}{\milli\watt} (orange). For the \qty{0}{\milli\watt} measurement, the lock-in amplifier time constant is \qty{200}{\micro\second}; for 1.8 and \qty{1.1}{\milli\watt} measurements, the lock-in time constant is \qty{100}{\micro\second}. (c) Total noise power for a noise-equivalent-power bandwidth of \qty{780}{\hertz}, measured in the frequency domain between \qty{204.8}{\kilo\hertz} to \qty{216.0}{\kilo\hertz} (red) and in the time domain (blue, shown in Fig. \ref{fig:spectroscopy}b). Error bars are 95\% confidence intervals.}
    \label{fig:noise}
\end{figure}

\newpage

\begin{figure}[h]
    \centering
    \makebox[\textwidth]{\includegraphics[width=13.5cm]{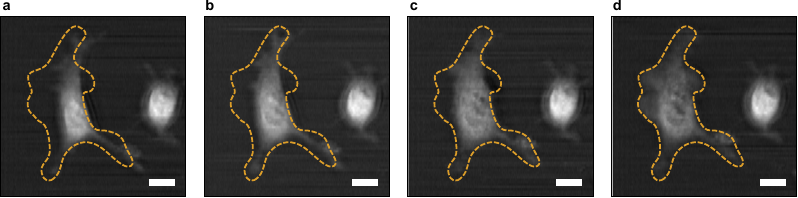}}
    \caption{\textbf{Extended Data Figure 4} Cavity-enhanced SRS microscopy of photodamage. A single position in the center of the left HeLa cell was exposed to the pump and Stokes beams continuously for \qty{30}{\second}, which initiated membrane blebbing and projection retraction, observed at times (a) 0 min, (b) 3 min, (c) 6 min, (d) 9 min at a Raman shift of \qty{2913}{\per\centi\meter}. The dashed, orange outline matches the cell shape in (d). The Stokes dose was \qty{14}{\milli\watt}. The pump dose was \qty{90}{\milli\watt}. Scale bars are \qty{10}{\micro\meter}.}
    \label{fig:blebbing}
\end{figure}

\newpage

\begin{figure}[h]
    \centering
    \includegraphics[width=8.45cm]{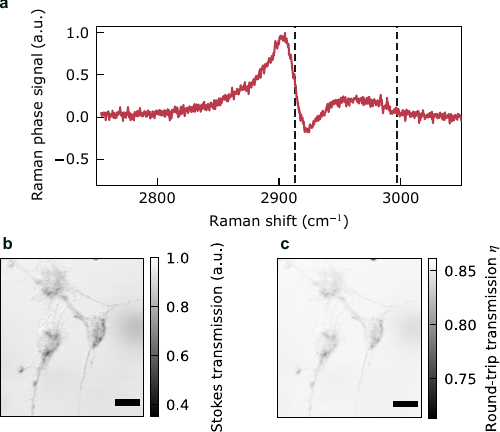}
    \caption{\textbf{Extended Data Figure 5} Additional measurement modalities. (a) Example SRS phase response of DMSO (red) measured with a pump wavelength of \qty{795}{\nano\meter}, which shows dispersive lineshapes around the Raman resonances at \qty{2913}{\per\centi\meter} and \qty{2997}{\per\centi\meter} (dashed, black). The asymmetry around the resonances is attributed to additional contributions from cross-phase modulation. (b) Stokes transmission image of RPE-1 cells recorded simultaneously with Fig. \ref{fig:hyper}. (c) Estimated $\eta$ from the cavity-enhanced Stokes transmission image. The contrast is reduced by 7.4x relative to (b). Scale bars are \qty{20}{\micro\meter}.} 
    \label{fig:phase}
\end{figure}

\end{appendices}

\end{document}